\documentclass[journal]{IEEEtran}

\ifCLASSINFOpdf
\else
   \usepackage[dvips]{graphicx}
\fi
\usepackage{url}

\usepackage{graphicx}
\usepackage{amsmath,amssymb,graphicx}
\usepackage[]{algorithm2e}
\usepackage{tabularx}
\usepackage{wasysym}
\usepackage{color,soul}
\usepackage{cite}
\usepackage{balance}
\usepackage{booktabs}
\usepackage{multirow}
\usepackage{siunitx}
\usepackage{hyperref}

\def\RR{{\mathbb R}}

\usepackage{makecell}
\usepackage{caption}
\usepackage[font=scriptsize]{subfig}
\newcommand{\subparagraph}{}
\usepackage{titlesec}
\usepackage{textgreek}

\usepackage{soul}
\soulregister\cite7 %
\soulregister\citep7 %
\soulregister\citet7 %
\soulregister\ref7 %
\soulregister\pageref7 %

\begin{document}

\setlength{\abovedisplayskip}{3.5pt}
\setlength{\belowdisplayskip}{3.5pt}

\titlespacing*{\section}{0pt}{2.ex plus 0ex minus 0ex}{0.5ex plus 0ex minus 0ex}
\titlespacing*{\subsection}{0pt}{2.ex plus 0ex minus 0ex}{0.5ex plus 0ex minus 0ex}

\title{ForkNet: Simultaneous Time and Time-Frequency Domain  Modeling for Speech Enhancement}

\author{Feng Dang, \IEEEmembership{Student Member, IEEE}, Qi Hu, Pengyuan Zhang, \IEEEmembership{Member, IEEE}, and Yonghong Yan, \IEEEmembership{Member, IEEE}
\thanks{Feng Dang is with the Institute of Information Engineering, Chinese Academy of Sciences, 100093, Beijing, China, and also with the School of Cyber Security, University of Chinese Academy of Sciences, 100049, Beijing, China. (e-mail: dangfeng19@mails.ucas.edu.cn).}
\thanks{Qi Hu, Pengyuan Zhang and Yonghong Yan are with the Key Laboratory of Speech Acoustics and Content Understanding, Institute of Acoustics, Chinese Academy of Science, Beijing, 100190, China. (email:\{huqi, zhangpengyuan, yanyonghong\}@hccl.ioa.ac.cn). (\textit{Corresponding authors: Qi Hu; Yonghong Yan.})}}

\markboth{Journal of \LaTeX\ Class Files, Vol. 14, No. 8, August 2015}
{Shell \MakeLowercase{\textit{et al.}}: Bare Demo of IEEEtran.cls for IEEE Journals}
\maketitle

\begin{abstract}
Previous research in speech enhancement has mostly focused on modeling time or time-frequency domain information alone, with little consideration given to the potential benefits of simultaneously modeling both domains. Since these domains contain complementary information, combining them may improve the performance of the model. In this letter, we propose a new approach to simultaneously model time and time-frequency domain information in a single model. We begin with the DPT-FSNet (causal version) model as a baseline and modify the encoder structure by  replacing the original encoder with three separate encoders, each dedicated to modeling time-domain, real-imaginary, and magnitude information, respectively. Additionally, we introduce a feature fusion module both before and after the dual-path processing blocks to better leverage information from the different domains.  The outcomes of our experiments reveal that the proposed approach achieves superior performance compared to existing state-of-the-art causal models, while preserving a relatively compact model size and low computational complexity.
The source code is available at \href{https://github.com/dangf15/ForkNet}{github.com/dangf15/ForkNet}.
\end{abstract}


\IEEEpeerreviewmaketitle

\section{Introduction}

\IEEEPARstart{S}{peech} enhancement (SE) is a pivotal technique in speech processing that endeavors to enhance the quality and intelligibility of speech that is corrupted by noise \cite {loizou2013speech}. This methodology is frequently utilized as a preliminary task for various applications such as automatic speech recognition, hearing aids, and telecommunications. In recent times, the adoption of deep neural networks (DNNs) in SE research has garnered significant attention from the scientific community.

Recent DNN-based methods have mainly focused on either the time domain using the encoder-separator-decoder scheme proposed in TasNet \cite{luo2018tasnet} or the complex time-frequency (T-F) domain using complex ratio masking. Both domains have unique advantages. The encoder and decoder of the time-domain models \cite{luo2019conv, chen20l_interspeech, luo2020dual} operate on short signal windows and have a higher potential upper bound due to the ability to train jointly with the backbone network. In contrast, complex T-F domain models \cite{tan2019learning, Hu2020, hao2021fullsubnet, li2021two, dang2022dpt, yu2022dbt, wang2022tf} utilize STFT features based on expert knowledge and typically have larger window sizes and hop sizes. A recent model called DBT-Net\cite{yu2022dbt} has been proposed that operates in the complex T-F domain and exhibits better performance in cross-corpus robustness than the representative time-domain models.

However,  the majority of previous models have predominantly focused on either time-domain or T-F domain modeling, with limited consideration of incorporating both types of information into a single model. Given that time and T-F domain information is complementary to some extent, integrating the strengths of both domains may lead to better overall performance. The most relevant models to our work are DTLN \cite{westhausen20_interspeech} and DBNet\cite{zhang21s_interspeech}.  DTLN combines the advantages of time-domain and T-F domain models, but uses two-stage networks to model T-F and time-domain information separately, effectively equivalent to connecting two different domain sub-models in series.  In DBNet, two branches are used to model time-domain and T-F domain information separately, with information exchange between these two branches facilitated by a single bridge branch, effectively equivalent to connecting two different domain sub-models in parallel. That is, DTLN and DBNet correspond to the time and T-F domain sub-models in series and parallel, respectively, and they do not exploit  both time and T-F domain information in a single model, thus potentially leading to sub-optimal performance. Therefore, it is a pertinent question to ask \textit{\textbf{whether it is possible to directly model both time-domain and frequency-domain information in a single model.}}


\begin{figure*}
  \centering  
  \includegraphics[width=\linewidth]{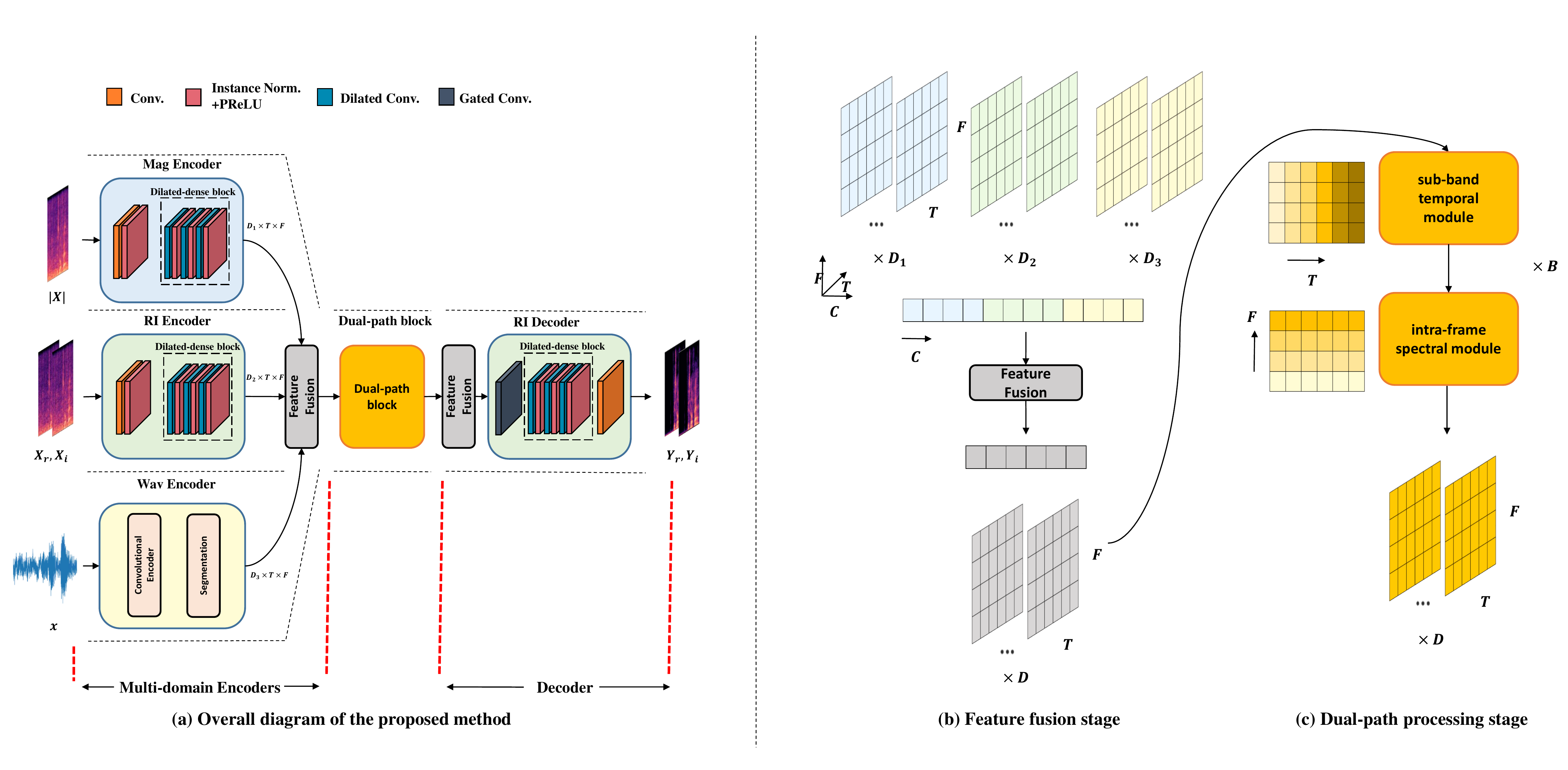}
  \vspace{-0.2cm}
  \caption{Architecture of the proposed ForkNet. (a) The overall diagram of the proposed method. (b) The detail of the feature fusion stage. (c) The detail of the dual-path processing stage.}
  \label{ovl}
  \vspace{-0.5cm}
\end{figure*}

The aforementioned discussion motivates us to propose a novel causal  SE framework that simultaneously models both time and time-frequency domain information in a single model. In this work, we make the following contributions:
\begin{itemize}
    \item   We adopt a powerful model, DPT-FSNet\cite{dang2022dpt}, as a baseline and modify it to a causal version to suit the task. To simultaneously model multi-domain information, we extend the original encoder structure into three separate encoders dedicated to modeling time-domain, real-imaginary (RI), and magnitude (Mag) information, respectively.
    \item  We introduce a feature fusion module both before and after the dual path processing module to enable it to better exploit information from different domains.
    \item  To evaluate the effectiveness of our proposed method, we conducted experiments on two benchmark datasets, namely the VoiceBank+DEMAND dataset\cite{valentini2016investigating} and the Interspeech 2020 Deep Noise Suppression (DNS) dataset\cite{ reddy20_interspeech}. Our findings indicate that our proposed model outperforms current state-of-the-art (SOTA) causal SE systems in terms of performance metrics, while maintaining  a smaller number of parameters.
\end{itemize}


\section{Proposed Algorithms}

 In SE, the mixed signal in the time domain can be formulated as ${x(n)}=s(n)+z(n)$, where $x(n),s(n),z(n)$ represent noisy speech, clean speech and noise, respectively. By applying the STFT, the signal in the time domain can be converted to the T-F domain, which can be expressed as:
\begin{footnotesize}
\begin{equation}
	{X_{f,t}}=S_{f,t}+Z_{f,t},
\end{equation}
\end{footnotesize}
where $X_{f,t}$, $S_{f,t}$, and $Z_{f,t}$ denote the complex spectral representations of the noisy, clean, and noise signals, respectively, in the frequency index $f$ and time frame index $t$.

 Fig.~\ref{ovl} shows the proposed system, dubbed ForkNet due to its fork-shaped structure.
To process a mixed signal $x$ in the time domain with shape $1 \times N$, we  first employ  multi-domain encoders to extract a high-dimensional feature with shape $D\times T\times F$. Specifically, each T-F unit of this high-dimensional feature is represented by a $D$-dimensional embedding that captures information from the complex and magnitude domains, as well as the time domain.
Subsequently, $B$ dual-path processing (DPP) blocks are utilized to refine the T-F embeddings. Each DPP block consists of a sub-band temporal module and an intra-frame spectral module to progressively exploit local and global spectro-temporal information.
Finally, an RI decoder is employed to obtain the estimated complex ratio mask, and inverse STFT (iSTFT) is utilized to re-synthesize the signal.
The rest of this section details the three components of ForkNet.

\subsection{Multi-domain Encoders}

Multi-domain encoders consist of three separate encoders for modeling time-domain, RI domain and Mag domain information respectively. Specifically, the RI and Mag encoders retain the same structure as the original encoder design. The Mag encoder has an input of $|X| \in \RR^{1\times T \times F}$ and an output of $H_{1} \in \RR^{D_{1}\times T \times F}$.  The RI encoder has an input of $X \in \RR^{2\times T \times F}$ and an output of $H_{2} \in \RR^{D_{2}\times T \times F}$. The time domain encoder consists of a convolutional encoder and a segmentation stage borrowed from DPRNN\cite{luo2020dual}. The input to the convolutional encoder is $x \in \RR^{1\times N}$ and the output is $W \in \RR^{D_{3}\times L}$, which is subsequently processed through the segmentation stage to obtain features of the same size as the RI and Mag encoders, $H_{3} \in \RR^{D_{3}\times T \times F}$.

The role of the feature fusion module (illustrated in Fig.~\ref{ovl}(b)) is to fuse information from the time, RI and Mag domains, enabling the DPP blocks to model multiple domains simultaneously.  The input $H \in \RR^{(D_{1}+D_{2}+D_{3})\times T \times F}$ of the feature fusion module is obtained by concatenating $H_1$, $H_2$, $H_3$ along the feature dimension and the output is a tensor $R_1 \in \RR^{D\times T \times F}$ containing multi-domain information.

\subsection{Dual-path Processing Block}

The dual-path processing block (illustrated in Fig.~\ref{ovl}(c)) consists of  a sub-band temporal module, and an intra-frame spectral module. In the sub-band temporal module, we view the input tensor $R_b \in \RR^{D\times T \times F}$ to the $b^{\text{th}}$ block as $F$ separate sequences, each with length $T$, and a causal GRU is used to model temporal information within each sub-band.
In the intra-frame spectral module,  the input tensor $U_b \in \RR^{D\times T\times F}$ is viewed as $T$ separate sequences, each with length $F$, and an improved transformer (the same as in \cite{dang2022dpt})  is used to model the  spectral information within each frame. The output tensor is denoted as $R_{b+1} \in \RR^{D\times T\times F}$.

\subsection{Decoder}
The RI decoder consists of a feature fusion block, a gated convolutional layer, a dilated dense block identical to that used in the RI encoder and a convolutional layer.
The feature $R_{B+1} $ from the final DPP block is passed through the decoder to obtain an estimated complex ratio mask. 
This mask is then used to perform element-wise multiplication with the RI encoder's input to obtain the enhanced complex spectrum $Y$. Finally, the enhanced speech waveform is obtained by passing the complex spectrum through the iSTFT.

\subsection{Multi-target Loss}
We train the ForkNet with the overall loss as follows:
\begin{equation}
  \mathcal{L} =  \mathcal{L}_\text{spec} + \lambda \mathcal{L}_\text{MR}
  \label{eq:mtloss}
\end{equation}
where $\lambda$  is the weighted coefficient between the two losses. The first loss, $\mathcal{L}_{spec}$, is the compressed spectral loss which is defined as follows:
\begin{small}
\begin{equation}
  \mathcal{L}_{spec} =
  \left \|\ |Y|^{c_{1}} - |S|^{c_{1}} \right \|_F^2 +  \left \|\ |Y|^{c_{1}} e^{j\varphi_Y} - |S|^{c_{1}} e^{j\varphi_S} \right \|_F^2
\end{equation}
\end{small}
where $c_{1}=0.6$ is a compression factor to model the perceived loudness \cite{Valin2020}. Moreover, the enhanced spectrogram $Y$ is converted to the time-domain and multiple STFTs are computed using windows ranging from \SIrange{5}{40}{\ms} in order to calculate the multi-resolution spectrogram loss:
\begin{small}
\begin{equation}
  \mathcal{L}_\text{MR} =
  \sum_{i} (\left \|\ |Y'_i|^{c_{2}} - |S'_i|^{c_{2}} \right \|_F^2  + \left \|\ |Y'_i|^{c_{2}} e^{j\varphi_Y} - |S'_i|^{c_{2}} e^{j\varphi_S} \right \|_F^2)
\end{equation}
\end{small}
where $Y'_i = \text{STFT}_i(y)$ is the i-th STFT of the predicted time-domain signal $y$ with windows sizes of $\{5, 10, 20, 40\}\si{\ms}$ respectively, and $c_{2}=0.3$ is a compression parameter. We empirically find that $\lambda = 1$ suffice in our evaluation.





\renewcommand\arraystretch{1.0}
\begin{table*}[t]
    \setcounter{table}{2}
	\caption{Comparisons with other state-of-the-art causal systems on the DNS Challenge dataset. ``-'' denotes no published result.}
    \small
	\centering
	\scalebox{0.85}{
		\begin{tabular}{cccccccccccc}
			\specialrule{0.1em}{0.25pt}{0.25pt}
			\multirow{2}*{Methods} &\multirow{2}*{Year} &\multirow{2}*{Params(M)}  &\multicolumn{4}{c}{w/ Reverberation} & \multicolumn{4}{c}{w/o Reverberation}\\
			\cmidrule(lr){4-7}\cmidrule(lr){8-11}
			& &  &WB-PESQ$\uparrow$ &PESQ$\uparrow$ &STOI(\%)$\uparrow$ &SISNR (dB)$\uparrow$ &WB-PESQ$\uparrow$ &PESQ$\uparrow$ &STOI(\%)$\uparrow$ &SISNR(dB)$\uparrow$\\
			\specialrule{0.1em}{0.25pt}{0.25pt}
			Noisy  &- &- &1.82 &2.75 &86.62 &9.03 &1.58 &2.45 &91.52 &9.07\\
			NSNet~{\cite{xia2020weighted}} &2020 &1.27 &2.37 &3.08 &90.43 &14.72 &2.15 &2.87  &94.47 &15.61\\
			DTLN~{\cite{westhausen20_interspeech}}  &2020 &0.99 &- &2.70 &84.68 &10.53 &- &3.04 &94.76  &16.34\\
			DCCRN~{\cite{Hu2020}} &2020 &3.67 &- &3.32 &- &- &- &3.27 &- &- \\
			FullSubNet~{\cite{hao2021fullsubnet}} &2021  &5.64 &2.97 &3.47 &92.62 &15.75 &2.78  &3.31 &96.11 &17.29\\
			CTS-Net~{\cite{li2021two}} &2021 &4.35 &3.02 &3.47 &92.70 &15.58 &2.94 &3.42  &96.66  &17.99\\
			FRNet~{\cite{li2022filtering}} &2022 &7.39 &3.18 &3.55 &93.25 &16.57 &3.17 &{3.56} &97.13 &18.91 \\
            CCFNet+~{\cite{dang2023first}} &2023 &0.62 &- &- &- &- &{3.06} &{3.59} &\textbf{97.26} &\textbf{19.27} \\
			\specialrule{0.1em}{0.25pt}{0.25pt}
            \textbf{ForkNet(Pro.)} &2023 &\textbf{0.58} &\textbf{3.27} &\textbf{3.71} &\textbf{94.12} &\textbf{16.79} &\textbf{3.17} &\textbf{3.62} &{97.16} &{19.18}\\
			\specialrule{0.1em}{0.25pt}{0.25pt}	
             
	\end{tabular}}
	\label{tbl:dns1}
\end{table*}
\vspace{0.2cm}

\renewcommand\arraystretch{1.0}
\begin{table*}[t]
\setcounter{table}{1}
\caption{Comparison with other state-of-the-art causal systems on the VoiceBank+DEMAND dataset.}
\small
\label{tab:1}
\centering

\begin{tabular}{c c c c c c c c c}
\specialrule{0.1em}{0.25pt}{0.25pt}	
 
Method & Feat. & Params(M)  & WB-PESQ  & STOI & CSIG & CBAK & COVL    \\
\specialrule{0.1em}{0.25pt}{0.25pt}	
Noisy  & - & - & 1.97  & 0.92 & 3.34 & 2.44 & 2.63   \\
\specialrule{0.1em}{0.25pt}{0.25pt}	
PercepNet~{\cite{Valin2020}} & Mag & 8.00 & 2.73  & - & - & - & -  \\
DEMUCS~{\cite{defossez20_interspeech}} & Waveform & 18.9  & 2.93  & 0.95 & 4.22 & 3.25 & 3.52  \\
DCCRN~{\cite{Hu2020}} & RI  & 3.70  & 2.54 & 0.94 & 3.74 & 3.13 & 2.75 \\
FullSubNet+~{\cite{chen2022fullsubnet+}} & RI  & 8.67   & 2.88  & 0.94 & 3.86 & 3.42 & 3.57  \\
CCFNet+~{\cite{dang2023first}} & RI  & 0.62 & 3.03  & 0.95 & 4.27 & 3.55 & 3.61   \\
CTS-Net~{\cite{li2021two}} &  RI + Mag   & 4.35 & 2.92  & - & 4.25 & 3.46 & 3.59  \\
DeepFilterNet2~{\cite{schroter2022deepfilternet2}} & RI + Mag  & {2.31} & {3.08}  & {0.94} & {4.30} & {3.40} & {3.70} \\
\specialrule{0.1em}{0.25pt}{0.25pt}	
\textbf{ForkNet(Pro.)} & RI + Mag + Waveform  & \textbf{0.58}   & \textbf{3.18}  & \textbf{0.95} & \textbf{4.39} & \textbf{3.65} & \textbf{3.81} \\
\specialrule{0.1em}{0.25pt}{0.25pt}	
\end{tabular}
\label{tbl:sota}
\vspace{-.4cm}
\end{table*}

\renewcommand\arraystretch{1.1}
\begin{table}[th]
\setcounter{table}{0}
\caption{Ablation analysis results on the VoiceBank+DEMAND dataset.}
\label{tab:ablation}
\scriptsize
\centering
\begin{tabular}{c c c c c c c c}
\toprule  
 
Method & Feat.  & Params.(M)  & WB-PESQ  & SI-SDR \\
\midrule
\textbf{Ref1}  & RI  & {0.76} & {3.11}  & {19.04} \\
\textbf{Ref2}  & RI + Mag  & {0.63} & {3.15}  & {19.11} \\
\textbf{ForkNet}  & RI + Mag + Waveform  & {0.58} & {3.18} & {19.22} \\ 
\bottomrule
\end{tabular}
\vspace{-.4cm}
\end{table}

\section{Experiments}

\subsection{Dataset}

In this work, we employ two datasets of differing scales to evaluate the performance of the proposed model. 
The first, referred to as the small-scale dataset, is the widely utilized VoiceBank+ DEMAND dataset \cite{valentini2016investigating} in SE research. The dataset comprises pre-mixed noisy speech and its corresponding clean speech segments, with the clean speech segments sourced from the VoiceBank corpus \cite{veaux2013voice}. The training set consists of 11,572 utterances from 28 speakers, and the test set includes 872 utterances from 2 speakers. During training, a noise set comprising 40 different noise conditions with 10 types of noises is employed, where 8 types of noises are selected from DEMAND \cite{thiemann2013diverse} and 2 types are artificially generated, with signal-to-noise ratios (SNRs) of 0, 5, 10, and 15 dB. For the test set, 20 different noise conditions with 5 types of unseen noise from the DEMAND database are utilized, with SNRs of 2.5, 7.5, 12.5, and 17.5 dB.

The second dataset, referred to as the large-scale dataset, is the DNS dataset \cite{reddy20_interspeech}. 
This dataset consists of a large amount of clean speech data that spans over 500 hours and has been collected from 2150 different speakers. Moreover, it also contains more than 180 hours of noise data, which have been collected from 150 different noise categories. To generate noisy-clean pairs during the training phase, we have used dynamic mixing.  Specifically, before the start of each training epoch,  $50\%$ of the clean speech segments are mixed with random room impulse responses (RIRs) provided by \cite{reddy2021icassp}. The speech-noise mixtures have been generated by mixing the clean speech segments ($50\%$ of which are reverberant) with noise at random SNRs between -5 and 20 dB. The DNS dataset includes two non-blind test sets named $\text{with\_reverb}$ and $\text{no\_reverb}$, both of which contain 150 noisy-clean pairs.

\subsection{Experimental Setup}
All the utterances are sampled at 16 kHz and chunked to 4 seconds to ensure training stability. 
For the STFT, the audio data were processed using a 32 ms Hanning window with a $50\%$ overlap of adjacent frames. Furthermore, 512 points Fast Fourier Transform (FFT) was employed, and the DC component of the spectrum was removed to generate 256-dimensional spectral features.

We use the configuration in \cite{dang2022dpt} to re-implement DPT-FSNet with some modifications to meet the causal settings and use it as a baseline, which is Ref1 as mentioned in Table~{\ref{tab:ablation}} below. We then replaced the RI encoder in Ref1 with three separate encoders; the Mag and RI encoders retain the same structure as the original encoder design, except that the number of channels is changed to $D_1$ and $D_2$ respectively, with both $D_1$ and $D_2$ set to 24.
The time-domain encoder consists of a convolutional encoder and a segmentation stage borrowed from DPRNN\cite{luo2020dual}, where the convolutional encoder has a window size of 2, a stride of 1 and a feature dimension of $D_3=16$.
The feature fusion module is implemented by a point-wise two-dimensional convolutional layer with a kernel size of 1x1. The first feature fusion module has $2D$ and $D$ input and output channels, respectively, and the second feature fusion module has $D$ and $2D$ input and output channels, respectively, where $D$ is set to 32. For more details, please refer to the model source code.

During the training phase, we trained the proposed models for 100 epochs, employing the Adam optimizer \cite{kingma2014adam}. To prevent gradient explosion, we applied gradient clipping with a maximum L2-norm of 5. The initial learning rate (LR) was 0.0004, which was subsequently reduced by a factor of 0.98 every two epochs.

\subsection{Evaluation Metrics}
We employ two objective evaluation metrics, namely wide-band PESQ (WB-PESQ) \cite{rec2005p} and short-time objective intelligibility (STOI) \cite{taal2011algorithm}, to assess the performance of our proposed model on both datasets. WB-PESQ measures the perceptual quality of the enhanced speech, while STOI evaluates the intelligibility of the speech signal.
In addition to WB-PESQ and STOI, we evaluate the VoiceBank+DEMAND dataset using three mean opinion score (MOS) predictors, namely CSIG, CBAK, and COVL \cite{hu2007evaluation}. The MOS predictors have a score range from 1 to 5.
For the DNS dataset, we also use scale-invariant signal-to-noise ratio (SI-SNR) and narrow-band PESQ (PESQ) as evaluation metrics to measure the performance of our model. Higher scores indicate better performance for all metrics.

\begin{figure*}
  \centering  
  \includegraphics[width=180mm]{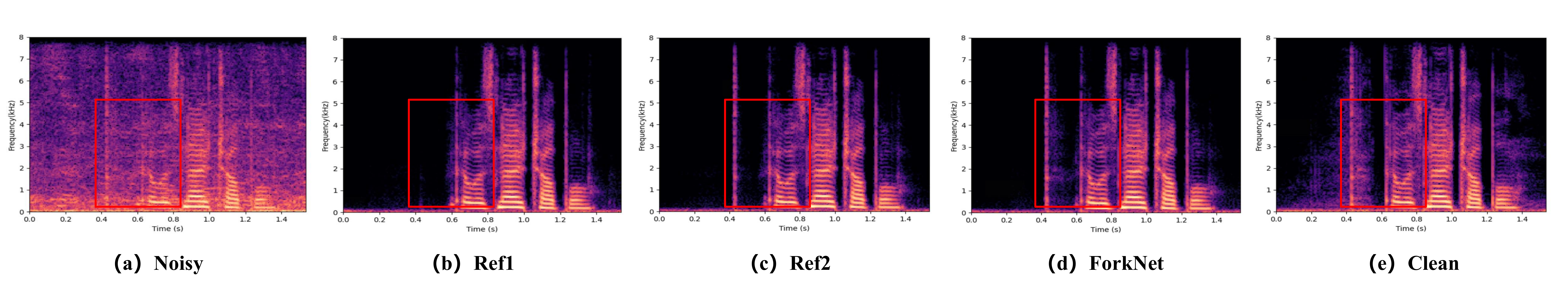}
  \caption{An illustration of the enhanced results using our proposed models. The speech spectrogram of (a) the noisy utterance. (b) the enhanced utterance by Ref1. (c) the enhanced utterance by Ref2. (d) the enhanced utterance by ForkNet. (e) the clean utterance.}
  \label{fig:abl}
  \vspace{-0.2cm}
\end{figure*}

\section{Evaluation Results}

\subsection{Ablation Analysis} 
In order to investigate the benefits of our time-domain and T-F domain simultaneous modeling strategy, we established two reference groups, denoted as Ref1 and Ref2. These groups, along with the proposed ForkNet model, vary solely in the encoder component. Ref1 employs a single  RI encoder that exclusively models RI information, with $D_1=0$, $D_2=2D$ and $D_3=0$. In contrast, Ref2 utilizes two independent encoders that model both RI and Mag information, with $D_1=D_2=D$ and $D_3=0$. To ensure a fair comparison, all three models satisfy the condition $\sum_{i} D_{i}=2D$. The comparison results among the three models are presented in Table~{\ref{tab:ablation}}. 
Two key observations can be made from our experiments. Firstly, the results demonstrate that moving from Ref1 to Ref2 yields consistent improvements, validating the effectiveness of supplementing Mag information with RI information. Secondly, ForkNet outperforms both Ref1 and Ref2 with fewer parameters, highlighting the advantages of our time-domain and T-F domain simultaneous modeling strategy. Fig.~{\ref{fig:abl}} validates our observations more intuitively.

\subsection{Comparison with Other SOTA Causal Methods}
Table~{\ref{tbl:sota}} presents the quantitative results of various advanced models operating in different domains on the VoiceBank+DEMAND dataset. It can be observed that the proposed ForkNet achieves overall SOTA performance with the least number of parameters, which verifies its superiority in terms of noise suppression and speech restoration. Additionally, Table~{\ref{tbl:dns1}} presents the results on the DNS non-blind test sets. Once again, the proposed system is highly competitive and outperforms other baselines in both the reverberation and anechoic cases with the least parameters, except for the anechoic case, where SI-SDR and STOI show slightly inferior results compared to CCFNet+. It should be noted that CCFNet+ only considered the case without reverberation.

\section{Conclusion}

We propose a simultaneous time-domain and time-frequency domain modeling approach for speech enhancement. It exploits the strengths of both domains, with improved performance in both baseline datasets. Notably, our approach achieves these gains without the need for additional parameters or computational effort compared to the complex T-F domain baselines. The proposed technique can be easily modified or directly adopted by numerous  model structures. Future research directions may include extending the proposed approach to multi-channel speech enhancement scenarios, where the simultaneous modeling of both domains may further improve the overall performance.



\bibliographystyle{IEEEtran}
\IEEEtriggeratref{17}
\bibliography{ref}
\end{document}